\documentclass[aps,prl,reprint,superscriptaddress]{revtex4-2}

\usepackage[english]{babel}
\usepackage[utf8]{inputenc}
\usepackage[top=2cm, bottom=2cm, left=2cm, right=2cm]{geometry}
\usepackage{color,soul}
\usepackage{graphicx}
\usepackage{siunitx}

\usepackage[svgnames]{xcolor}
\usepackage{amsmath}
\usepackage{tabulary}

\usepackage{tikz}

\usepackage[euler]{textgreek}

\begin{document}

\title{\AA ngstrom depth resolution with chemical specificity at the liquid-vapor interface}

\author{R.~Dupuy}
\email{remi.dupuy@sorbonne-universite.fr}
\affiliation{Fritz-Haber-Institut der Max-Planck-Gesellschaft, Faradayweg 4-6, 14195 Berlin, Germany}
\affiliation{Sorbonne Université, CNRS, Laboratoire de Chimie Physique - Matière et Rayonnement, LCPMR, F-75005 Paris Cedex 05, France}
\author{J.~Filser}
\affiliation{Fritz-Haber-Institut der Max-Planck-Gesellschaft, Faradayweg 4-6, 14195 Berlin, Germany}
\author{C.~Richter}
\affiliation{Fritz-Haber-Institut der Max-Planck-Gesellschaft, Faradayweg 4-6, 14195 Berlin, Germany}
\author{T.~Buttersack}
\affiliation{Fritz-Haber-Institut der Max-Planck-Gesellschaft, Faradayweg 4-6, 14195 Berlin, Germany}
\author{F.~Trinter}
\affiliation{Fritz-Haber-Institut der Max-Planck-Gesellschaft, Faradayweg 4-6, 14195 Berlin, Germany}
\affiliation{Institut für Kernphysik, Goethe-Universität Frankfurt am Main, Max-von-Laue-Str.\ 1, 60438 Frankfurt am Main, Germany}
\author{S.~Gholami}
\affiliation{Fritz-Haber-Institut der Max-Planck-Gesellschaft, Faradayweg 4-6, 14195 Berlin, Germany}
\author{R.~Seidel}
\affiliation{Helmholtz-Zentrum Berlin für Materialien und Energie, Albert-Einstein-Str.\ 15, 12489 Berlin, Germany}
\author{C.~Nicolas}
\affiliation{Synchrotron SOLEIL, L'Orme des Merisiers, Saint-Aubin - BP 48 91192, Gif-sur-Yvette Cedex, France}
\author{J.~Bozek}
\affiliation{Synchrotron SOLEIL, L'Orme des Merisiers, Saint-Aubin - BP 48 91192, Gif-sur-Yvette Cedex, France}
\author{D.~Egger}
\affiliation{Fritz-Haber-Institut der Max-Planck-Gesellschaft, Faradayweg 4-6, 14195 Berlin, Germany}
\author{H.~Oberhofer}
\affiliation{Department of Physics, University of Bayreuth, 95440 Bayreuth, Germany}
\author{S.~Th\"urmer}
\affiliation{Department of Chemistry, Graduate School of Science, Kyoto University, Kitashirakawa-Oiwakecho, Sakyo-Ku, Kyoto 606-8502, Japan}
\author{U.~Hergenhahn}
\affiliation{Fritz-Haber-Institut der Max-Planck-Gesellschaft, Faradayweg 4-6, 14195 Berlin, Germany}
\author{K.~Reuter}
\email{reuter@fhi.mpg.de}
\affiliation{Fritz-Haber-Institut der Max-Planck-Gesellschaft, Faradayweg 4-6, 14195 Berlin, Germany}
\author{B.~Winter}
\affiliation{Fritz-Haber-Institut der Max-Planck-Gesellschaft, Faradayweg 4-6, 14195 Berlin, Germany}
\author{H.~Bluhm}
\email{bluhm@fhi.mpg.de}
\affiliation{Fritz-Haber-Institut der Max-Planck-Gesellschaft, Faradayweg 4-6, 14195 Berlin, Germany}

\begin{abstract}

The determination of depth profiles across interfaces is of primary importance in many scientific and technological areas. Photoemission spectroscopy is in principle well suited for this purpose, yet a quantitative implementation for investigations of liquid-vapor interfaces is hindered by the lack of understanding of electron-scattering processes in liquids. Previous studies have shown, however, that core-level photoelectron angular distributions (PADs) are altered by depth-dependent elastic electron scattering and can, thus, reveal information on the depth distribution of species across the interface. Here, we explore this concept further and show that the experimental anisotropy parameter characterizing the PAD scales linearly with the average distance of atoms along the surface normal obtained by molecular dynamics simulations. This behavior can be accounted for in the low-collision-number regime. We also show that results for different atomic species can be compared on the same length scale. We demonstrate that atoms separated by about 1~\AA~along the surface normal can be clearly distinguished with this method, achieving excellent depth resolution.

\end{abstract}

\maketitle

Photoemission spectroscopy (PES) has become an important technique for the investigation of liquid-vapor interfaces \cite{dupuy2021b,ammann2018,winter2006,brown2009a}, especially since the advent of liquid microjets. PES provides fundamental, molecular-level information on these interfaces. For instance, the electronic energetics of solvated species can be determined \cite{seidel2016,tang2010a}, allowing for an accurate determination of ionization energies as well as work-function measurements \cite{thurmer2021,tissot2016,perezramirez2021}. Ultrafast processes, e.g., non-local decay pathways \cite{jahnke2020} or electron delocalization \cite{nordlund2007} after excitation were also investigated. Basic yet important questions in chemistry, such as the surface propensity and surface protonation state of solutes can be addressed. More challenging measurements, for instance the investigation of heterogeneous reactions with trace gases are also now becoming possible \cite{artiglia2017}.

An important property of the liquid-vapor interface is the depth distribution and concentration of surfactants and dissolved species in the interfacial region, which can significantly differ from that in the bulk. Depth profiling via PES can be obtained through varying the escape depth of the photoelectrons, either by changing their take-off angle relative to the surface normal (which is only possible under specific experimental conditions, not met, in particular, by cylindrical microjets) or the photoelectron kinetic energy (eKE) by changing the incident photon energy. The eKE influences the inelastic mean free path (IMFP) of the electrons, i.e., the mean distance electrons travel between inelastic-scattering events. Inelastic scattering results in a loss of kinetic energy of the photoelectrons (in this context a loss of several eV) and hence their removal from the photoelectron signal of interest. The IMFP is, thus, related (in a non-straightforward way) to the escape depth.

The dependence of PE intensities on the escape depth has been widely used in liquid-phase PES \cite{holmberg1986,ghosal2005} to reveal the depth distribution of species at the interface. It has, however, been hampered by several issues. First and foremost, exact values of the IMFP and escape depths in liquid water are still debated \cite{signorell2020,shinotsuka2017,ottosson2010,suzuki2014,thurmer2013a}, in particular at low eKE. Even at higher eKE, the uncertainty is still too large to allow for a reliable calculation of depth profiles from PE intensities \cite{dupuy2021b}. Another under-explored quantity is the photoionization cross section of the relevant core levels of the solutes, which relies on calculated atomic data. Ionization cross sections are, for instance, known to exhibit oscillations in molecular, including condensed-phase, systems \cite{bjorneholm2014}. For these reasons, obtaining precise and reliable depth profiles from eKE-dependent measurements is often not possible.

Alternative PES-based depth-profiling techniques for the solid state are peak-shape analysis \cite{tougaard2010} or X-ray standing wave \cite{zegenhagen2013} techniques, which are, however, not suitable for liquids and also have their own drawbacks. A few other techniques exist to obtain molecular-scale depth profiles of liquid-vapor interfaces. For example, X-ray reflectivity (XRR) probes the electron density along the surface normal with a resolution down to 3~\AA, and when used resonantly can achieve some degree of elemental specificity \cite{bu2015}. Its application to liquid-vapor interfaces is well established \cite{bera2018}. Another less common technique is neutral-ion backscattering \cite{andersson2020}, which can similarly achieve a 2--3~\AA~resolution depth profile, although it requires complex data analysis to extract element-specific profiles. These two techniques nonetheless lack the chemical specificity of PES.

Depth profiling based on photoelectron angular distributions (PADs) from core levels has been explored in a few previous studies \cite{lewis2019,dupuy2022}. In the condensed phase, the nascent (intrinsic molecular) PAD is modified by elastic electron scattering, which reduces the inherent anisotropy by randomizing the electron trajectories. The PAD for randomly oriented molecules and linearly polarized X-rays is described by \cite{reid2003}

\begin{equation}
    \mathrm{PAD}(\theta) \sim 1 + \frac{\beta}{2}(3\cos^2(\theta)-1)
\end{equation}

where $\theta$ is the angle between the linear polarization vector of the incident beam and the electron emission direction, and $\beta$ is the so-called anisotropy parameter, characterizing the angular distribution of photoelectron emission, ranging from $\beta = 2$ to -1. At the so-called magic angle (54.7$^{\circ}$), photoemission becomes independent of \textbeta. Elastic electron scattering leads to a reduction of the measured $\beta$ value as compared to that of the nascent distribution, $\beta_{nasc}$. Since the average number of elastic collisions encountered increases with increasing travel distance to the surface, a given reduction in anisotropy can, in principle, be assigned to a given distance to the surface of the point of origin of the photoelectrons.

In this letter, we show how measured PADs can be used to distinguish the average relative depth of atoms within a molecule with a precision of 1~\AA. We also show that the relationship between the anisotropy parameter and the probing distance along the surface normal is linear, at least near the interface, and explain this behavior.

Experiments were performed at two different liquid-jet setups installed respectively at the PLEIADES beamline (SOLEIL) and the UE52\_SGM beamline (BESSY~II) \cite{seidel2017}. At both beamlines, the linear polarization vector of the soft X-rays can be freely varied to form an angle between $0^\circ$ and $90^\circ$ with respect to the measurement direction, allowing to measure PADs without reorienting the setup, detector, or sample. More experimental details are available in the Supplementary Information (SI) and in previous works \cite{dupuy2022,thurmer2021,gozem2020,seidel2017}.

\begin{figure}
    \includegraphics[trim={0cm 0cm 12cm 0cm},clip,width=\linewidth]{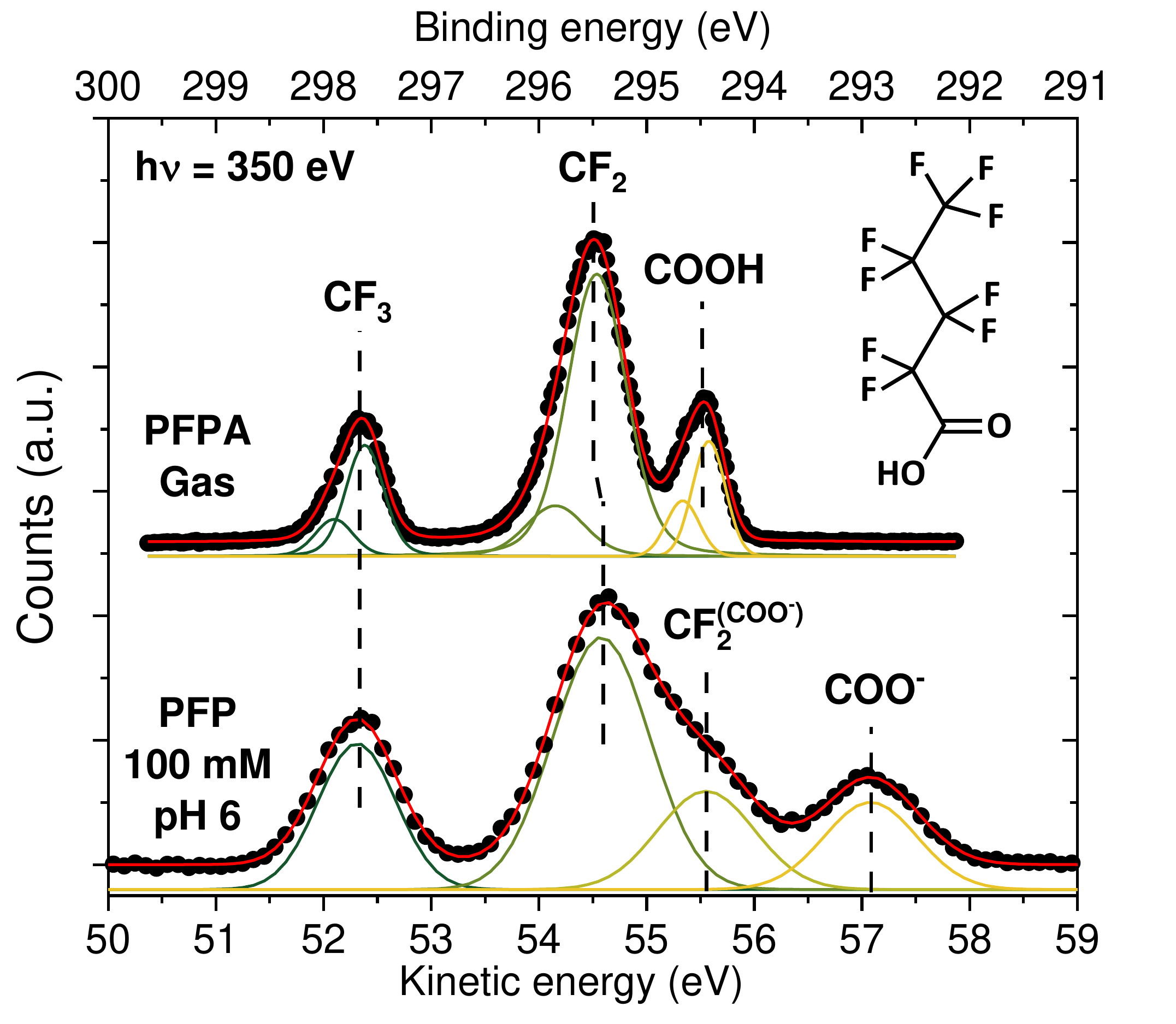}
    \caption{C~1s PE spectra of gaseous perfluorinated pentanoic acid (PFPA) and of an aqueous solution of 100~mM perfluoropentanoate (PFP) at pH 6 measured with a photon energy of 350~eV, corresponding to a kinetic energy of $\sim$50-60~eV, and at the magic angle. Peak assignments are labeled. Both spectra are aligned on the binding-energy axis at the CF$_3$ peak position; the binding energy axis is not calibrated. Gas-phase peaks were fitted using two gaussians for each peak, to account for vibrational asymmetry.}
    \label{PFPA_spectra}
\end{figure}

The present study is devoted to perfluorinated pentanoic acid (PFPA) surfactants on an aqueous solution. Perfluorinated surfactants are widely used in industry \cite{lemal2004}, but are very persistent pollutants and have become a major environmental concern \cite{lau2012}. For our purpose, perfluorinated carboxylic acids possess favorable spectroscopic properties compared to regular hydrogenated molecules. Owing to the large electronegativity of the fluorine atom, the CF$_3$ and CF$_2$ carbons are easily distinguished in PES, and both peaks are also well separated from the COOH carbon peak, as can be seen in the C~1s gas-phase spectrum of PFPA in Fig.~\ref{PFPA_spectra}. This makes it possible to distinguish both ends of the molecule. When in solution, the molecule dissociates (the pKa is close to 0 \cite{cabala2017}) to its deprotonated form (perfluoropentanoate or PFP). PFP exhibits a fourth distinguishable carbon peak in the C~1s PES spectrum (Fig.~\ref{PFPA_spectra}), since the COO$^-$ peak is shifted to lower binding energy compared to COOH, as expected, and a shoulder at the low-binding-energy side of the CF$_2$ peak can be attributed to the nearest-neighbor CF$_2$ carbon of the COO$^-$ group. The molecule can, thus, be probed at four different sites along the molecular axis.

We also performed measurements of the O~1s level, where the peak of the oxygen atoms from COO$^-$ can be distinguished from that of liquid and gas-phase water. O~1s spectra together with additional spectra are shown in the SI. Since we expect the surfactant molecules to accumulate at the interface and orient with the hydrophilic COO$^-$ group towards the interface while the hydrophobic perfluorinated tail points towards the vacuum, we are able to probe atoms at different well-defined distances along the surface normal. These assumptions are confirmed with the help of the molecular dynamics (MD) simulations that will be presented below.

First, we need to measure the nascent $\beta_{nasc}$ of the molecule in the gas phase, before considering the anisotropy reduction in the condensed phase. For closed, purely atomic core s-orbitals a value of $\beta = 2$ is expected at all eKEs. However, intramolecular scattering and possible changes in orbital character already modify the anisotropy of the PAD of the isolated molecule in the gas phase. To quantify these effects, we performed gas-phase PAD measurements, which we consider to represent the nascent distributions of the liquid-phase molecule. This approximation is discussed in the SI. We have already observed previously \cite{dupuy2022} that $\beta_{gas}$ can be different in value for different functional groups of a given molecule. Here, we measured PADs at three different eKEs, 50~eV, 150~eV, and 350~eV. $\beta_{gas}$ is found to be different for the CF$_3$, CF$_2$, and COOH carbons at all eKEs (values are given in the SI), highlighting again the necessity of these gas-phase measurements for proper analysis of liquid-phase results. For the remainder of this letter, we will consider the reduction factor in anisotropy from gas to liquid, R$_{\beta}$~=~$\beta_{liq}$/$\beta_{gas}$, as the relevant quantity instead of raw $\beta$ values. For instance, R$_{\beta}$~=~0.8 corresponds to a 20\% reduction in anisotropy of the nascent PAD caused by elastic scattering. All measured raw values of $\beta_{gas}$ and $\beta_{liq}$ are tabulated in the SI.

\begin{figure}
    \includegraphics{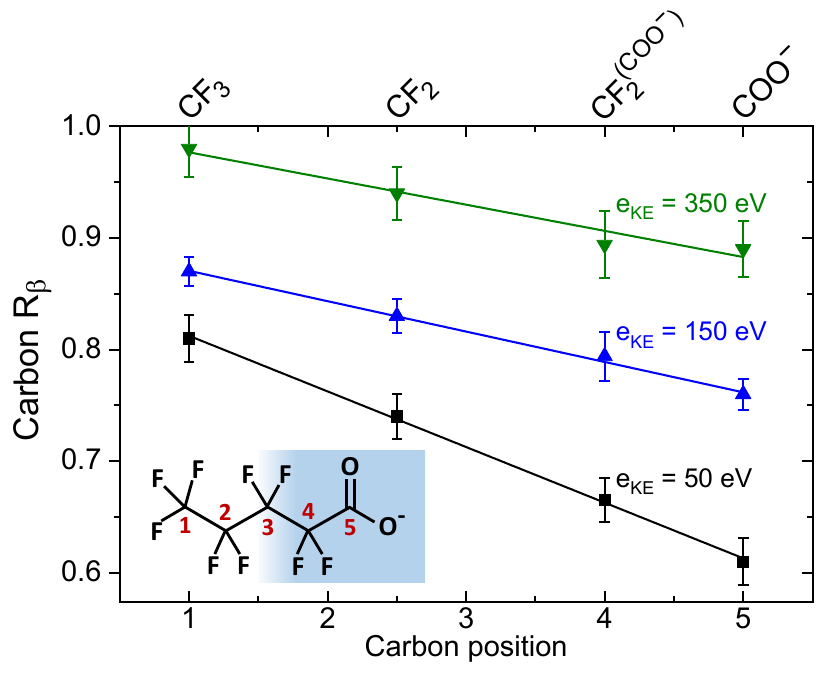}
    \caption{Relative anisotropy parameter $R_{\beta}$ of the different carbon sites for a 100~mM PFP solution at pH 6, measured at different eKEs. Results are given for carbon positions 1 to 5 indicated in the structural formula of PFP in the inset, which also shows the expected interface arrangement of the molecule with respect to the water surface. Since carbon 2 and 3 cannot be resolved, the corresponding data point is plotted at position 2.5. A linear trend was fitted to the data.}
    \label{beta_eKE}
\end{figure}

R$_{\beta}$ values for a 100~mM PFP solution at different eKEs are displayed in Fig.~\ref{beta_eKE}. R$_{\beta}$ decreases progressively from the CF$_3$ carbon to the COO$^-$ group carbon, which is the expected behavior for a molecule standing up relative to the surface, with the COO$^-$ group (partly) solvated and the CF$_3$ pointing towards vacuum. We observe that R$_{\beta}$ decreases linearly with carbon position for all eKEs. If we assume equal spacing along the surface normal between all carbons, this means that there is a linear relationship between the average position along the surface normal of the carbon atoms and R$_{\beta}$.

Before addressing this point in more detail, let us further discuss the results of Fig.~\ref{beta_eKE}. One can observe that the slope for the eKE~=~50~eV data is higher than that for eKE~=~150~eV, which in turn is slightly higher than that for eKE~=~350~eV. This behavior is indeed qualitatively expected, since the slope should mainly depend on the number of elastic-scattering events per unit of distance, and thus on the inelastic to elastic mean free path ratio, which increases with decreasing eKE.

Another notable finding is the behavior of R$_{\beta}$ for the CF$_{3}$ group. In the very simple picture sketched in the inset of Fig.~\ref{beta_eKE}, it may seem surprising that even the outermost carbon atom experiences scattering (intramolecular scattering being already taken into account by the normalization). However, due to the cylindrical geometry of the jet, the signal is sampled over a wide range of electron take-off angles relative to the surface normal. Furthermore, some disorder is inherent to the liquid surface, where molecules are not necessarily standing straight up and the surface is not atomically flat. Electrons may, thus, scatter on neighboring surfactant molecules even when they originate from the CF$_3$ group.

\begin{figure}
    \includegraphics[trim={0cm 0cm 0cm 0cm},clip,width=\linewidth]{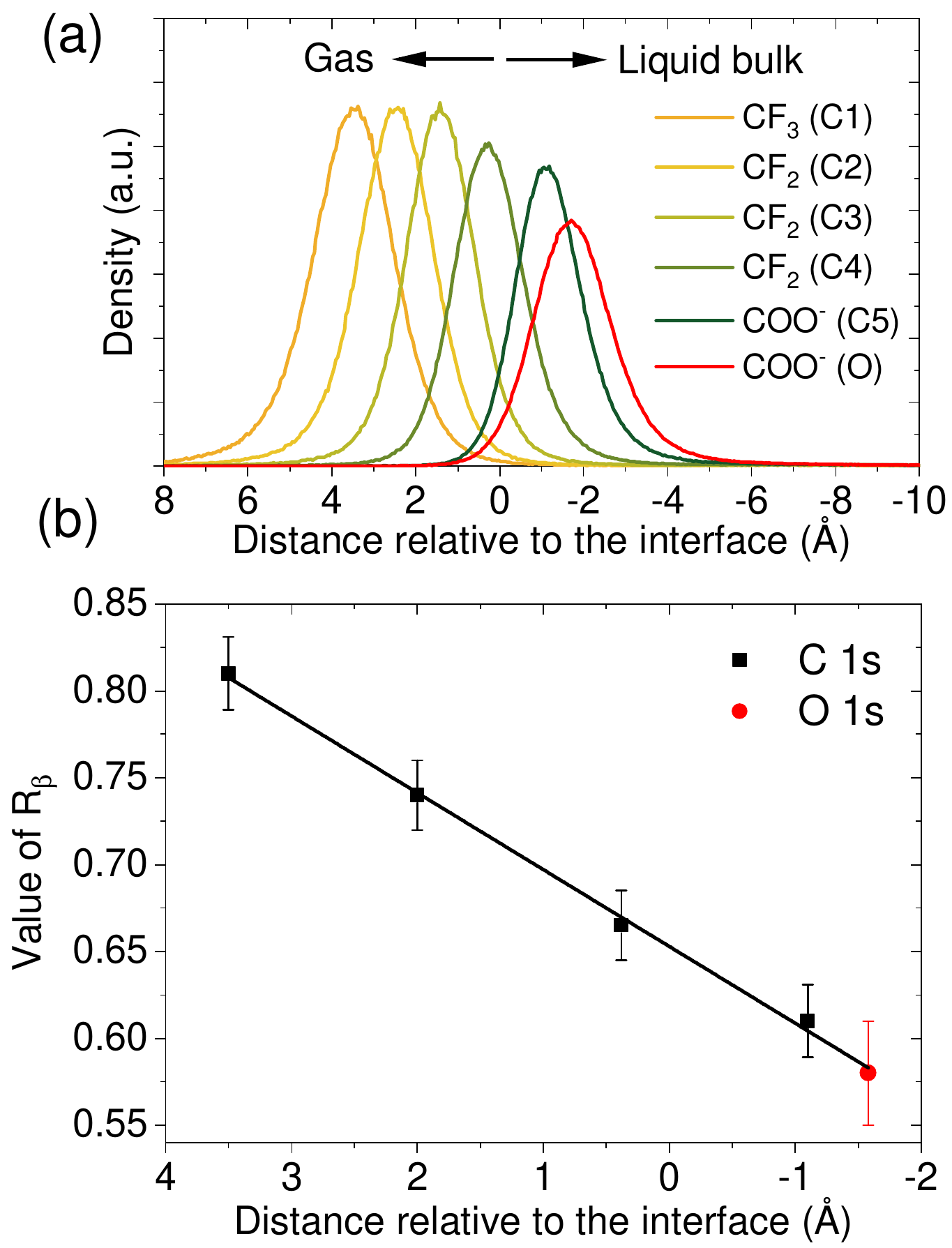}
    \caption{(a) MD-simulations-derived atomic distributions as a function of the distance to the water surface, modified by an exponential attenuation with characteristic length of 12~\AA~(see the text for details). The position of z = 0 is defined in the simulations according to the instantaneous interface (see ref. \cite{willard2010} and SI). (b) Anisotropy parameter R$_{\beta}$ (normalized to the gas phase) of the C~1s carbon peaks and COO$^-$ O~1s peak of the PFP molecule, as a function of the average atomic distance to the surface obtained from panel (a).}
    \label{beta_all}
\end{figure}

To gain more insights into the interface arrangement of the perfluorinated surfactant, we performed MD simulations of NaPFP on a water slab. Details can be found in the SI \cite{PLIMPTON19951,lammps,lammps_paper2022,SPC_Fw,OPLS,LigParGen,CM5,CM5-calculator,hirshfeld,pbe0,aims,Ren_2012,YU2018267,HAVU20098367,Ihrig_2015,mbd-nl,Na_FF,pppm_textbook,pppm_disp,fftool,packmol,Nose,Hoover,watkins2001,ase-paper,l2reg}.
Vertical distributions of the C and O atoms along the $z$ axis (i.e., the global surface normal) were determined relative to the instantaneous surface of the polar phase \cite{Liu2018}. Figure~\ref{beta_all}(a) displays these distributions, modified by an exponential attenuation factor to account for the fact that the measured signal is exponentially attenuated with depth. We chose an effective attenuation length (EAL) of 12~\AA, a reasonable value at 50~eV eKE for water \cite{thurmer2013a}, as an approximation. This exponential attenuation does not significantly change the distribution amplitudes as the size of PFP is small compared to the EAL. The non-attenuated distributions are shown in Fig.~S5 in the SI.

Figure~\ref{beta_all}(b) displays the same R$_{\beta}$ parameters at the C~1s edge for eKE~=~50~eV reported in Fig.~\ref{beta_eKE}, but with the $x$ axis now corresponding to the (EAL-modified) average distance relative to the interface derived from Fig.~\ref{beta_all}(a). The value for the COO$^-$ oxygen atoms from the O~1s data is included as well. The linear decrease of R$_{\beta}$ with increasing distance to the interface is confirmed and the R$_{\beta}$ value for the COO$^-$ oxygen atoms aligns well with the linear trend for the carbon atoms. Figure~\ref{beta_all} demonstrates that PADs of different core levels from different atomic species align on the same scale, and confirms the linear trend for R$_{\beta}$ as a function of distance to the interface.

The sensitivity of the measurement derived from the slope in Fig.~\ref{beta_all} is $\sim$0.045 per \AA. Full error bars, on the other hand, range between 0.04 and 0.06 per \AA. Repeated measurements at eKE $\sim$150~eV (see Fig.~\ref{beta_eKE}) over multiple campaigns were stable within these uncertainties (see SI). The resolution achieved in this experiment is therefore of the order of 1~\AA~or less under favorable conditions. Indeed, for all (classes of) carbon atoms R$_{\beta}$ can be well distinguished above the error bars, and the carbon atoms are separated on average by 1.5~\AA~along the surface normal.

The sensitivity will vary depending on the details of the PAD measurements, such as the probed core level and the incident photon energy. For instance, the slope of R$_{\beta}$ and thus the sensitivity of the measurement depends on eKE (see Fig.~\ref{beta_eKE}). Ultimately, it will depend on the elastic-scattering properties of the system. The achievable sensitivity will also be affected by the precision with which R$_{\beta}$ values can be determined.

We now come back to the linearity of the R$_{\beta}$ scale and discuss its physical origin. To understand this behavior, we need to consider how the PAD is affected by elastic scattering. The differential cross section (DCS) for elastic scattering, which gives the probability of a deviation from the original trajectory by an angle $\theta$ in a single scattering event, is what relates elastic scattering to anisotropy reduction. For an average number of elastic collisions $n$, Th\"urmer et al.\ in their work on the O~1s PADs of liquid water expressed the modified distribution I*(\texttheta) as the $n$-fold convolution of the initial distribution I(\texttheta) with the DCS: I*(\texttheta)~=~I(\texttheta)$\ast$(DCS(\texttheta))$^n$ \cite{thurmer2013a}. This one-dimensional approach ignores specific geometric effects such as the finite angular acceptance of the electron analyzer, but it is sufficient for our purpose. Th\"urmer et al. showed that the DCS for water can be reasonably approximated by a simple Gaussian of the form exp($-\theta^2/2\phi^2$), with a `width' of $\phi$~=~17$^{\circ}$, reproducing the measured gas-phase DCS for water satisfactorily. We will adopt the same approximation here, i.e., evaluate the modification of the nascent PAD after $n$ average elastic collisions by calculating the convolution product given above for a Gaussian DCS. One can then deduce the anisotropy parameter $\beta^*$ of the new distribution as a function of the initial $\beta$:

\begin{equation}
\label{betastar}
    \beta^* = \beta\frac{\mathrm{e}^{-2n\phi^2}}{1 + \frac{\beta}{4}(1-\mathrm{e}^{-2n\phi^2})}
\end{equation}

For a low number of collisions (n$\to$0), Eq.~\ref{betastar} reduces to:

\begin{equation}
\label{betastar_dl}
    \frac{\beta^*}{\beta} = 1 - (1 + \frac{\beta}{4})2n\phi^2
\end{equation}

The difference between the exact formulation (Eq.~\ref{betastar}) and a linear approximation (Eq.~\ref{betastar_dl}) for $\phi$~=~17$^{\circ}$ is 2\% for $n = 1$, 9\% for $n = 2$, and only reaches a significant value of 26\% for $n = 3$. Thus, we can consider that for an average number of elastic collisions $n < 2$, $\beta^*$ scales linearly with $n$. For an atom located at a specific depth $z$ along the surface normal, $n$ is directly proportional to $z$. From this simple model, we can account for the observed linear behavior of the measured $R_{\beta}$ as a function of $z$.

While Eqs.~\ref{betastar} and \ref{betastar_dl} were calculated for the Gaussian approximation of the DCS, they turn out to be valid for other DCS, with only the effective value of $\phi$ changing. This is discussed in the SI. There are, however, other limitations of this model, such as the fact that the interface exhibits a complex structure with potentially different scattering properties in the surfactant layer below and above the average water level. In fact, one could expect a different behavior above and below $z = 0$, as atoms are no longer surrounded by water molecules. Above $z = 0$, contributions to scattering only include backscattering on and within the water surface and scattering on other neighboring surfactant molecules. However, as far as can be inferred from the limited number of points and the error bars, no variation in the slope is observed at $z = 0$, which may imply that the scattering properties in water and within the PFP surfactant layer are not sufficiently different to observe a marked change. We note again that the cylindrical geometry of the jet combined with the relative disorder of the surfactant layer likely averages out any specific geometric effect.

While it is not possible to derive quantitative information from the slopes observed in Fig.~\ref{beta_eKE}, we can qualitatively account for the observed linear relationship between R$_{\beta}$ and the average distance to the interface, which is the important conclusion here. Indeed, at a photoelectron kinetic energy of 50~eV, we can assume an elastic MFP on the order of 5--8~\AA~for both water and the surfactant, although the exact value is not known. Within the PFP layer, elastic electron scattering should, thus, remain in the single-scattering regime $(n = 1)$, where linearity is expected. On the other hand, for electrons experiencing many scattering events, i.e., in case of large elastic-scattering cross sections and/or atoms deep within the interface, the linear relationship should break down.

In conclusion, we have demonstrated a promising depth-profiling technique based on core-level PAD measurements, and made the following observations: (i) The average distance of the origin of a set of photoelectrons to the interface relates linearly with the reduction of its $\beta$ parameter relative to the gas phase, R$_{\beta}$. (ii) A common R$_{\beta}$ scale can be established for all atomic species based on measurements of different core levels. (iii) \AA-scale resolution can be achieved, i.e., distinguishable PE peaks originating at an average depth difference of 1~\AA~can exhibit a difference in R$_{\beta}$ exceeding the one-standard-deviation error in this quantity. The sensitivity depends on the experimental conditions.

Compared with other methods mentioned in the introduction of this letter, core-level PAD depth profiling achieves better depth resolution and is not only element-specific but also sensitive to the chemical environment. On the other hand, it only accesses the average depth and not the full depth profile. Nonetheless, our results show that core-level PAD measurements are a powerful tool to study liquid interfaces. The technique should also be applicable to solid-vacuum or solid-vapor interfaces of similarly amorphous systems.

\begin{acknowledgments}

We acknowledge Synchrotron SOLEIL for provision of synchrotron radiation at the PLEIADES beamline, under projects 20201076 and 20211393, and also HZB (Helmholtz-Zentrum Berlin) for the provision of synchrotron radiation at BESSY~II, beamline UE52\_SGM, under project 202-09662. We thank the technical service personnel of the SOLEIL chemistry laboratories for their helpful support. R.D. acknowledges support from the Alexander von Humboldt foundation through a Postdoctoral Fellowship. B.W., U.H., and T.B. acknowledge support from the European Research Council (883759-AQUACHIRAL). F.T. and B.W. acknowledge support by the MaxWater initiative of the Max-Planck-Gesellschaft. S.T. acknowledges support from the JSPS KAKENHI Grant No. JP20K15229.

\end{acknowledgments}

\newpage
\foreach \x in {1,...,12}
{%
\clearpage
\includegraphics[page={\x},scale=0.9,trim={2cm 1.5cm 0cm 2cm},clip]{Supp_Mat.pdf}
}

\end{document}